\documentclass[12pt]{iopart}
\usepackage{graphicx}
\usepackage[english]{babel}
\usepackage{bm}
\usepackage{iopams}

\begin{document}
\title[Vacuum magnetic birefringence: PVLAS]{Measurements of vacuum magnetic birefringence using permanent dipole magnets: the PVLAS experiment}

\author{F Della Valle$^1$, U Gastaldi$^2$, G Messineo$^3$, E Milotti$^1$, R Pengo$^2$, L Piemontese$^3$, G Ruoso$^2$ and G Zavattini$^3$}
\address{$^1$ INFN, Sez. di Trieste and Dip. di Fisica, Universit\`a di Trieste, via A. Valerio 2, I-34127 Trieste, Italy}
\address{$^2$ INFN, Lab. Naz. di Legnaro, viale dell'Universit\`a 2, I-35020 Legnaro, Italy}
\address{$^3$ INFN, Sez. di Ferrara and Dip. di Fisica, Universit\`a di Ferrara, via Saragat 1, Edificio C, I-44122 Ferrara, Italy}
\ead{federico.dellavalle@ts.infn.it}

\begin{abstract}

The PVLAS collaboration is presently assembling a new apparatus (at the INFN section of Ferrara, Italy) to detect vacuum magnetic birefringence (VMB). VMB is related to the structure of the QED vacuum and is predicted by the Euler-Heisenberg-Weisskopf effective Lagrangian. It can be detected by measuring the ellipticity acquired by a linearly polarised light beam propagating through a strong magnetic field. Using the very same optical technique it is also possible to search for hypothetical low-mass particles interacting with two photons, such as axion-like (ALP) or millicharged particles (MCP). Here we report results of a scaled-down test setup and describe the new PVLAS apparatus. This latter one is in construction and is based on a high-sensitivity ellipsometer with a high-finesse Fabry-Perot cavity ($>4\times 10^5$) and two 0.8~m long 2.5~T rotating permanent dipole magnets. Measurements with the test setup have improved by a factor 2 the previous upper bound on the parameter $A_e$, which determines the strength of the nonlinear terms in the QED Lagrangian: $A_e^{\rm (PVLAS)} < 3.3 \times 10^{-21}$~T$^{-2}$ 95\% c.l. Furthermore, new laboratory limits have been put on the inverse coupling constant of ALPs to two photons and confirmation of previous limits on the fractional charge of millicharged particles is given.
\end{abstract}

\pacs{07.60.Fs, 12.20.-m, 42.50.Xa, 78.20.Ls}
\submitto{\NJP}
\noindent{\it magneto-optical tests of QED; nonlinear electrodynamics; PVLAS\/}
\maketitle

\section{Introduction}	

In the absence of matter, Maxwell's equations can be obtained from the classical electromagnetic Lagrangian density ${\cal L}_{\rm Cl}$ (in S.I. units)
 \begin{equation}
 {\cal L}_{\rm Cl} = \frac{1}{2\mu_{\rm 0}}\left(\frac{E^{2}}{c^{2}}-B^{2}\right)
 \end{equation}
where $\mu_{0}$ is the magnetic permeability of vacuum and $c$ is the speed of light in vacuum.
A quadratic Lagrangian leads to linear partial differential equations for the fields, and the superposition principle holds, thereby excluding light-light scattering and other non linear electromagnetic effects in vacuum.

With the introduction of Dirac's equation for electrons and Heisenberg's Uncertainty Principle, Euler, Heisenberg and Weisskopf in 1936 \cite{QED} derived an effective Lagrangian density which leads to electromagnetic non linear effects even in vacuum. For photon energies well below the electron mass and fields much smaller than their critical values, $B \ll B_{\rm crit} = {m_{e}^{2}c^{2}}/{e \hbar} = 4.4\times10^{9}$~T, $E \ll E_{\rm crit} = {m_{e}^{2}c^{3}}/{e \hbar} = 1.3\times10^{18}$~V/m, the Euler-Heisenberg-Weisskopf effective Lagrangian can be written as
 \begin{equation}
 {\cal L}_{\rm EHW} = {\cal L}_{\rm Cl}+\frac{A_{e}}{\mu_{\rm 0}}\Bigg[\Big(\frac{E^{2}}{c^{2}}-B^{2}\Big)^{2}+7\Big(\frac{\vec{E}}{c}\cdot\vec{B}\Big)^{2}\Bigg]
 \label{LEH}
 \end{equation}
where
 \begin{equation}
 A_{e} = \frac{2}{45\mu_{0}}\frac{\alpha^{2}\mathchar'26\mkern-10mu\lambda_e^{3}}{m_{e}c^{2}} = 1.32\times10^{-24}~{\rm T}^{-2}
 \label{Ae}
 \end{equation}
with $\mathchar'26\mkern-10mu\lambda_e=\hbar/mc$ being the Compton wavelength of the electron, $\alpha={e^2}/{(\hbar c 4\pi\epsilon_0)}$ the fine structure constant, $m_e$ the electron mass. 

\begin{figure}[htb]
\begin{center}
\includegraphics[width=10cm]{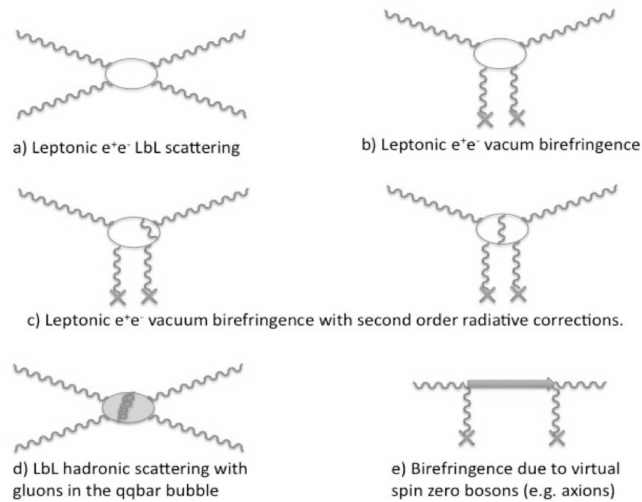}
\end{center}
\caption{Feynman diagrams for four field interactions.}
\label{4photon}
\end{figure}

This Euler-Heisenberg-Weisskopf Lagrangian allows four field interactions that can be represented, to first order, by the Feynman diagrams shown in Figure \ref{4photon} a) and b).
Figure \ref{4photon} a) represents light by light scattering whereas Figure \ref{4photon} b) represents the interaction of real photons with a classical field leading to vacuum magnetic birefringence.

Also Quantum Electrodynamics (QED) predicts non-linear effects in vacuum leading to birefringence and light-by-light (LbL) scattering through the four-photons box diagram \cite{Schwinger1951,Adler,Iacopini1979,DeTollis1964,Karplus1951,Dicus1998,Bernard2000,Moulin1996}. Furthermore, magnetic birefringence (see Figure \ref{4photon} e)) and magnetic dichroism could be generated by hypothetical axion-like particles (ALPs) \cite{Maiani1986,Sikivie1983,Gasperini1987,Raffelt1988} or millicharged particles (MCPs) \cite{Ahlers2007,Gies2006,Tsai1975,Daugherty1983,Schubert2000}. Finally, the coupling of four photons through $q{\bar q}$ fluctuations is also possible (see Figure \ref{4photon} d)). In view of the values of the masses of even the lightest quarks (much heavier than the electron mass) this last contribution should be very small. However, theoretical predictions have a very large uncertainty -- they span three orders of magnitude \cite{Rafelski1998}; moreover, it is not possible to evaluate the QCD contributions from indirect measurements, unlike what happened for the muon $g-2$ \cite{Davier2009,Nyffeler2010}.

Vacuum magnetic linear birefringence and light-light interaction in vacuum at very low energies have yet to be observed. Several experimental efforts are underway to detect such effects \cite{PVLAS1998,PRD2008,Lundstrom2006,exawatt2008,Luiten2004,BMV2008,Ni1996,Pugnat2005,heinzl2006}. Before the present work, the previous bound on four photon interaction was set by the PVLAS collaboration \cite{Bregant2008} with an upper bound on vacuum magnetic birefringence $\Delta n^{\rm (PVLAS)}$ measured with a magnetic field $B=2.3$~T
\begin{equation}
\Delta n^{\rm (PVLAS)}<1.0\times10^{-19}\qquad@\;2.3~{\rm T}~{\rm and~95\%~c.l.}
\end{equation}
The measurement has been performed at $\lambda=1064$~nm, corresponding to a photon energy $\hbar\omega=1.17$~eV. The limit translates in an upper bound for unpolarised photon-photon interaction \cite{DeTollis1964,Karplus1951,Dicus1998,Bernard2000,Moulin1996} as
\begin{equation}
\sigma_{\gamma\gamma}^{\rm (PVLAS)}(1.17~{\rm eV})< 4.6\times10^{-62}\;{\rm m}^{2}.
\end{equation}
The predicted QED value of the vacuum magnetic birefringence (see below) $\Delta n^{\rm (EHW)}$ and photon-photon elastic scattering cross section $\sigma_{\gamma \gamma}^{\rm (EHW)}$ are
 \begin{eqnarray}
 \label{Deltan}
 \Delta n^{\rm (EHW)} &=& 2.1\times10^{-23}\qquad @ \; 2.3~{\rm T}\\
 \sigma_{\gamma\gamma}^{\rm (EHW)}({\rm 1.17~eV}) &=& 1.8\times10^{-69}~{\rm m}^{2}
 \end{eqnarray}

\subsection{Electrodynamics}
 
To calculate the magnetic birefringence of vacuum one can proceed by determining the electric displacement vector $\vec D$ and the magnetic intensity vector $\vec H$ from the Lagrangian density $\cal L_{\rm EHW}$ of Eq. (\ref{LEH}) by using the constitutive relations \cite{Adler}
 \begin{equation}
 \vec D = \frac{\partial {\cal L}}{\partial \vec E} \quad {\rm and} \quad \vec H = - \frac{\partial {\cal L}}{\partial \vec B}.
 \end{equation}
 From these one obtains
 \begin{eqnarray}
 \vec D &=&\epsilon_{\rm 0}\vec E + \epsilon_{0}A_{e}\Big[4\Big(\frac{E^{2}}{c^{2}}-B^{2}\Big)\vec E + 14 \Big(\vec E\cdot\vec B\Big)\vec B\Big] \label{D}\\
 \vec H &=& \frac{\vec B}{\mu_{\rm 0}} + \frac{A_{e}}{\mu_{\rm 0}}\Big[4\Big(\frac{E^{2}}{c^{2}}-B^{2}\Big)\vec B - 14 \bigg(\frac{\vec E}{c}\cdot\vec B\bigg)\frac{\vec E}{c}\Big].
 \label{H}
 \end{eqnarray} 
Let's consider a linearly polarised beam of light propagating perpendicularly to an external magnetic field $\vec B_{\rm ext}$. With $\vec D$ and $\vec H$, and using Maxwell's equations in media one can describe light propagation in an external field. It is evident that the equations for light propagation will no longer be linear due to the non linear dependence of $\vec D$ and $\vec H$ on $\vec E$ and $\vec B$, respectively. We substitute $\vec E = \vec E_{\gamma}$ and $\vec B = \vec B_{\gamma} + \vec B_{\rm ext}$ in (\ref{D}) and (\ref{H}), where the index $\gamma$ refers to the light, assuming $\vec B_{\rm ext}\gg B_\gamma$. One finds the following relations for the relative dielectric constant and magnetic permeability of vacuum:
 \begin{eqnarray}
 \left\{ \begin{array}{ll}
 \epsilon_{\parallel}^{\rm (EHW)} &= 1 + 10 A_{e}B_{\rm ext}^2\\
 \mu_{\parallel}^{\rm (EHW)} &= 1 + 4 A_{e}B_{\rm ext}^2\\
 n_{\parallel}^{\rm (EHW)} &= 1 + 7 A_{e}B_{\rm ext}^2
 \end{array} \right.
\quad
 \left\{ \begin{array}{ll}
 \epsilon_{\perp}^{\rm (EHW)} &= 1 - 4 A_{e}B_{\rm ext}^2\\
 \mu_{\perp}^{\rm (EHW)} &= 1 + 12 A_{e}B_{\rm ext}^2\\
 n_{\perp}^{\rm (EHW)} &= 1 + 4 A_{e}B_{\rm ext}^2
 \end{array} \right.
 \label{index}
 \end{eqnarray}
where the indices $\parallel$ and $\perp$ refer to  light polarisation parallel and perpendicular to $\vec B_{\rm ext}$, respectively.

From these sets of equations two important consequences are apparent: the velocity of light in the presence of an external magnetic field is no longer $c$ and vacuum is birefringent with
 \begin{equation}
 \Delta n^{\rm (EHW)}=n_{\parallel}^{\rm (EHW)}-n_{\perp}^{\rm (EHW)} = 3 A_{e} B_{\rm ext}^2.
 \label{birifQED}
 \end{equation}
Numerically this leads to the value given in equation (\ref{Deltan}). In any respect, magnetised vacuum behaves like a uniaxial crystal.
 
\subsection{Post-Maxwellian generalisation}

It is interesting to generalise the non linear electrodynamic Lagrangian density correction, which must be parity conserving, gauge and Lorentz invariant, by introducing two free parameters $\eta_{\rm 1}$ and $\eta_{\rm 2}$:
 \begin{equation}
 {\cal L}_{\rm pM} = \frac{\xi}{2 \mu_{\rm 0}}\left[\eta_{\rm 1}\left(\frac{E^{2}}{c^{2}}-B^{2}\right)^{2}+4\eta_{\rm 2}\left(\frac{\vec{E}}{c}\cdot\vec{B}\right)^{2}\right]
 \end{equation}
where $\xi = 1/B_{\rm crit}^{2}$, 
and $\eta_{\rm 1}$ and $\eta_{\rm 2}$ are dimensionless parameters depending on the model. With such a formulation the birefringence induced by an external transverse magnetic field is
 \begin{equation}
 \Delta n^{\rm (pM)} = 2 \xi\left(\eta_{\rm 2}-\eta_{\rm 1}\right) B_{\rm ext}^2.
 \end{equation}
This expression reduces to Eq. (\ref{birifQED}) if $\eta_{\rm 1} = \alpha /45 \pi$ and $\eta_{\rm 2} = 7/4 \eta_{\rm 1}$. It is apparent that $n_{\parallel}$ depends only on $\eta_{1}$ whereas $n_{\perp}$ depends only on $\eta_{2}$. It is also noteworthy that if $\eta_{\rm 1} = \eta_{\rm 2}$, as is the case in the Born-Infeld model \cite{BornInfeld}, then there is no magnetically induced birefringence even though elastic scattering is present \cite{Haissinsky2006,Denisov2004}.

\subsection{Higher order QED corrections}

Figure \ref{4photon} c) shows Feynman diagrams for the $\alpha^{3}$ contribution to the vacuum magnetic birefringence. The effective Lagrangian density for this correction has been evaluated by different authors \cite{Ritus1975,bakalovrad,Dunne2004} and can be expressed as
\begin{equation}
{\cal L}_{\rm rad} = \frac{A_{e}}{\mu_{0}}\left(\frac{\alpha}{\pi}\right)\frac{10}{72}\left[32\Big(\frac{E^{2}}{c^{2}}-B^{2}\Big)^{2}+263\Big(\frac{\vec{E}}{c}\cdot\vec{B}\Big)^{2}\right].
\end{equation}
This Lagrangian leads to an extra correction $\Delta n^{\rm (rad)}$ to the vacuum magnetic birefringence given in equation (\ref{birifQED})
\begin{equation}
\Delta n^{\rm (rad)}=\frac{25\alpha}{4\pi}3 A_{e}B_{\rm ext}^{2}=0.0145 \times 3 A_{e}B_{\rm ext}^{2}
\end{equation}
resulting in a 1.45 \% correction to $\Delta n^{\rm (EHW)}$.

\subsection{Other hypothetical effects}
 
\label{sec:3}
Two other important hypothetical effects could cause $n\ne1$ and $\Delta n \ne 0$ in the presence of an external magnetic field transverse to the light propagation direction. They can be due either to neutral bosons weakly coupled to two photons, like the axion \cite{Maiani1986,Sikivie1983,Gasperini1987,Raffelt1988} or to millicharged particles (MCP) \cite{Ahlers2007,Gies2006,Tsai1975,Daugherty1983,Schubert2000}.

\subsubsection{Axion-like particles (ALP's)}

Search for axions or axion-like particles using laboratory optical techniques was experimentally pioneered by the BFRT collaboration \cite{Cameron1993} and subsequently continued by the PVLAS collaboration with an apparatus installed at INFN National Laboratories in Legnaro (LNL) \cite{PVLAS1998,PRD2008,Bregant2008}. In 2006 the detection of a dichroism induced by a magnetic field in vacuum was published \cite{PRL2006}. Such a result, although in contrast with model-dependent interpretations of the results of the CAST experiment \cite{CAST}, could have been due to axion-like particles. Soon after, this result was excluded by the same PVLAS collaboration \cite{PRD2008,Bregant2008} after a series of upgrades to the apparatus, and almost simultaneously the axion-like interpretation was excluded by two other groups \cite{RizzoALP1,RizzoALP2,gammeV} with regeneration-type measurements.

The Lagrangian densities describing the interaction of axion-like particles with two photons, for convenience expressed in natural Heaviside-Lorentz units, can be written as
\begin{equation}
{\cal L}_{a} = g_a\phi_{a} \vec{E}\cdot \vec{B} \quad{\rm and}\quad
{\cal L}_{s} = g_s\phi_{s} \left(E^{2}-B^{2} \right)
\label{lagalp}
\end{equation}
where $g_a$ and $g_s$ are the coupling constants of the pseudoscalar field $\phi_{\rm a}$ or scalar field $\phi_{\rm s}$, respectively.
Therefore, in the presence of an external uniform magnetic field $\vec{B}_{\rm ext}$ the component of the electric field of light $\vec{E}_{\rm \gamma}$ parallel to $\vec{B}_{\rm ext}$ will interact with the pseudoscalar field. For the scalar case the opposite is true: an interaction is only possible for the component of $\vec{E}_{\rm \gamma}$ normal to $\vec{B}_{\rm ext}$.

For photon energies above the mass $m_{a,s}$ of such particle candidates, a real production can follow: the oscillation of photons into such particles decreases  the intensity of only one of the polarisation components. On the other hand, even if the photon energy is smaller than the particle mass, virtual production can take place, causing a phase delay of one component with respect to the other. The phase difference $\Delta\varphi=\phi_{\parallel}-\phi_{\perp}$ accumulated in an optical path $L$ inside the magnetic field region will generate a birefringence
\begin{equation}
\Delta n = \frac{\Delta\varphi}{2\pi}\frac{\lambda}{L}.
\end{equation}

We use the complex index of refraction, written as $\tilde{n}=n+i\kappa$, where $n$ is the index of refraction and $\kappa$ is the extinction coefficient, which is related to the absorption coefficient $\mu$ by $\mu=4\pi\kappa/\lambda$. A dichroism can be described by the difference in extinction coefficient $\Delta\kappa=\kappa_\parallel-\kappa_\perp$ of the medium for the two orthogonal polarisations. In the pseudoscalar case, where $n^{a}_{\parallel}>1$, $\kappa^a_\parallel>0$, $n^{a}_{\perp}=1$ and $\kappa^a_\perp=0$, and in the scalar case, where $n^{s}_{\perp}>1$, $\kappa^s_\perp>0$, $n^{s}_{\parallel}=1$ and $\kappa^s_\parallel=0$, the dichroism $\Delta \kappa$ and the birefringence $\Delta n$ can be expressed, in both the scalar and pseudoscalar cases, as \cite{Maiani1986,Sikivie1983,Cameron1993}:
\begin{eqnarray}
&|\Delta\kappa|=\displaystyle\kappa_{\parallel}^{a} = \kappa_{\perp}^{s} = 2\left(\frac{g_{a,s}B_{\rm ext}L}{4}\right)^{2}\left(\frac{\sin x}{x}\right)^{2}\\
&|\Delta n|=\displaystyle n_{\parallel}^{a}-1 = n_{\perp}^{s}-1 = \frac{g_{a,s}^2B_{\rm ext}^{2}}{2m_{a,s}^{2}}\left(1-\frac{\sin2x}{2x}\right)
\label{pseudo}
\end{eqnarray}
where, in vacuum, $x=\frac{Lm_{a,s}^{2}}{4\omega}$, $\omega$ is the photon energy and $L$ is the magnetic field length. The above expressions are in natural Heaviside-Lorentz units so that 1~T $=\sqrt{\frac{\hbar^{3}c^{3}}{e^{4}\mu_{0}}}= 195$~eV$^2$ and 1~m $=\frac{e}{\hbar c}=5.06\times10^{6}$~eV$^{-1}$. 

In the approximation for which $x\ll1$ (small masses) expression (\ref{pseudo}) becomes
\begin{equation}
|\Delta n| = n^{a}_{\parallel}-1 = n^{s}_{\perp}-1 = \frac{g_{a,s}^2B_{\rm ext}^{2}m_{a,s}^{2}L^{2}}{16}
\end{equation}
whereas for $x \gg 1$
\begin{equation}
|\Delta n| = n^{a}_{\parallel}-1 = n^{s}_{\perp}-1 = \frac{g_{a,s}^2B_{\rm ext}^{2}}{2m_{a,s}^{2}}.
\end{equation}
It is interesting to note that the birefringences induced by pseudoscalars and scalars are opposite in sign: $n_{\parallel}^a > n_{\perp}^a = 1$ whereas $n_{\parallel}^s = 1 < n_{\perp}^s$. The different behavior of $n^{s}_{\perp}-1$ and $n^{a}_{\parallel}-1$ with respect to $L$ in the two cases where $x\ll1$ and $x\gg1$ leaves in principle, if the magnet length can be varied, a free experimental handle for distinguishing between the two scenarios: this is in fact the case of the new PVLAS setup.

\subsubsection{MCP}

Consider now the vacuum fluctuations of particles with charge $\pm\epsilon e$ and mass $m_{\epsilon}$ as discussed in references \cite{Ahlers2007} and \cite{Gies2006}. The photons traversing a uniform magnetic field may interact with such fluctuations, resulting in both a pair production and a phase delay if the photon energy $\omega > 2m_{\epsilon}$, and only a phase delay if $\omega < 2m_{\epsilon}$. We are discussing here only the birefringence and hence the (real) index of refraction. Moreover, we consider separately the case of fermions and of bosons.

 - Dirac fermions

Let us first consider the case in which the millicharged particles are Dirac fermions (Df). As derived by Tsai and Erber in reference \cite{Tsai1975}, the indices of refraction of photons with polarisation respectively parallel and perpendicular to the external magnetic field have two different mass regimes defined by a dimensionless parameter $\chi$ (S.I. units):
\begin{equation}
\chi\equiv
\frac{3}{2}\frac{\hbar\omega}{m_{\epsilon}c^{2}}\frac{\epsilon e B_{\rm ext}\hbar}{m_{\epsilon}^{2}c^{2}}.
\label{chi}
\end{equation}
It can be shown that \cite{Ahlers2007,Daugherty1983}
\begin{equation}
n_{\parallel,\perp}^{\rm Df}=1+I_{\parallel,\perp}^{\rm Df}(\chi) A_{\epsilon} B_{\rm ext}^{2}
\label{ndf}
\end{equation}
with
\begin{equation}
I_{\parallel,\perp}^{\rm Df}(\chi)=
\left\{ \begin{array}{ll}
\left[\left(7\right)_{\parallel},\left(4\right)_{\perp}\right] & \textrm { for  } \chi \ll 1 \\
\displaystyle-\frac{9}{7}\frac{45}{2}\frac{\pi^{1/2}2^{1/3}\left(\Gamma\left(\frac{2}{3}\right)\right)^{2}}{\Gamma\left(\frac{1}{6}\right)}\chi^{-4/3}\left[\left(3\right)_{\parallel},\left(2\right)_{\perp}\right] & \textrm{ for   }\chi\gg 1
\end{array}\right.\nonumber
\label{idf}
 \end{equation}
and
\begin{equation}
A_{\epsilon}=\frac{2}{45\mu_{0}}\frac{\epsilon^{4}\alpha^{2} \mathchar'26\mkern-10mu\lambda_\epsilon^{3}}{m_{\epsilon}c^{2}}
\end{equation}
in analogy to equation (\ref{Ae}).
In the limit of large masses ($\chi\ll1$) this expression reduces to (\ref{index}) with the substitution of $\epsilon e$ with $e$ and $m_{\epsilon}$ with $m_{e}$ in equation (\ref{ndf}). The dependence on $B_{\rm ext}$ remains the same as for the well known QED prediction.

For small masses ($\chi\gg1$) the index of refraction now also depends on the parameter $\chi^{-4/3}$ resulting in a net dependence of $n$ with $B_{\rm ext}^{2/3}$ rather than $B_{\rm ext}^{2}$. 

In both mass regimes, a birefringence is induced:
\begin{eqnarray}
\label{deltandf}
&\Delta n^{\rm Df}&=\left[I_{\parallel}^{\rm Df}(\chi)-I_{\perp}^{\rm Df}(\chi)\right] A_{\epsilon} B_{\rm ext}^{2}=\\
&=&\left\{\begin{array}{ll}
3 A_{\epsilon} B_{\rm ext}^{2}& \textrm{ for  } \chi \ll 1 \\
\displaystyle-\frac{9}{7}\frac{45}{2}\frac{\pi^{1/2}2^{1/3}\left(\Gamma\left(\frac{2}{3}\right)\right)^{2}}{\Gamma\left(\frac{1}{6}\right)}\chi^{-4/3}A_{\epsilon} B_{\rm ext}^{2}& \textrm{ for   }\chi\gg 1.
\end{array}\right.\nonumber
\end{eqnarray}

 - Scalar particles
 
 Very similar expressions to the Dirac fermion case can also be obtained for the scalar (sc) charged particle case \cite{Ahlers2007,Schubert2000}. Again there are two mass regimes defined by the same parameter $\chi$ of expression (\ref{chi}). In this case the indices of refraction for the two polarisation states with respect to the magnetic field direction are
 \begin{equation}
n_{\parallel,\perp}^{\rm sc}=1+I_{\parallel,\perp}^{\rm sc}(\chi) A_{\epsilon} B_{\rm ext}^{2}
\label{nsc}
 \end{equation}
 with
 \begin{equation}
I_{\parallel,\perp}^{\rm sc}(\chi)
=\left\{ \begin{array}{ll}
\left[\left(\frac{1}{4}\right)_{\parallel},\left(\frac{7}{4}\right)_{\perp}\right] & \textrm { for  } \chi \ll 1 \\
\displaystyle-\frac{9}{14}\frac{45}{2}\frac{\pi^{1/2}2^{1/3}\left(\Gamma\left(\frac{2}{3}\right)\right)^{2}}{\Gamma\left(\frac{1}{6}\right)}\chi^{-4/3}\left[\left(\frac{1}{2}\right)_{\parallel},\left(\frac{3}{2}\right)_{\perp}\right] & \textrm{ for   }\chi\gg 1.
\end{array}\right.\nonumber
\label{isc}
 \end{equation}
 
The magnetic birefringence is therefore
\begin{eqnarray}
\label{deltansc}
&\Delta n^{\rm sc}&=\left[I_{\parallel}^{\rm sc}(\chi)-I_{\perp}^{\rm sc}(\chi)\right] A_{\epsilon} B_{\rm ext}^{2}=\\
&=&\left\{\begin{array}{ll}
\displaystyle-\frac{6}{4} A_{\epsilon} B_{\rm ext}^{2}& \textrm{ for  } \chi \ll 1 \\
\displaystyle\frac{9}{14}\frac{45}{2}\frac{\pi^{1/2}2^{1/3}\left(\Gamma\left(\frac{2}{3}\right)\right)^{2}}{\Gamma\left(\frac{1}{6}\right)}\chi^{-4/3}A_{\epsilon} B_{\rm ext}^{2}& \textrm{ for   }\chi\gg 1.
\end{array}\right.\nonumber
\end{eqnarray} 
As can be seen, there is a sign difference in the birefringence $\Delta n$ induced by an external magnetic field in the presence of Dirac fermions with respect to the case in which scalar particles exist.

\section{Experimental method}


A birefringence $\Delta n$ induces an ellipticity $\psi$ on a linearly polarised beam of light given by
\begin{equation}
\psi(\vartheta) = \frac{\Delta\varphi}{2}\sin2\vartheta= \pi \frac{L\Delta n}{\lambda}\sin2\vartheta
\label{Psi}
\end{equation}
where $L$ is the optical path length within the birefringent region and $\lambda$ is the wavelength of the light traversing it; $\vartheta$ is the angle between the light polarisation and the slow axis, defined in our case by the magnetic field direction. The optical path length can be increased by using a Fabry-Perot cavity; in this case we think of an effective path length $L_{\rm eff}$ and use a capital letter $\Psi$ to indicate the total induced ellipticity. In fact, given a birefringent region of length $L$ within a Fabry-Perot cavity of finesse $\cal F$, the effective path length is $L_{\rm eff} = 2 {\cal F}L/\pi$ (see section C below). Today finesses ${\cal F} > 400000$ can be obtained.

In the Euler-Heisenberg-Weisskopf case, $\Delta n$ depends quadratically on the magnetic field $\vec{B}$. High magnetic fields can be obtained with superconducting magnets but, as we will see below, it is desirable to have a time dependent field either by modulating its intensity, thereby changing $\Delta n$ or by rotating the field direction, thereby changing $\vartheta$. This makes superconducting magnets far less appealing than permanent magnets which, today, can reach fields above 2.5~T. In the BFRT experiment the first scheme was adopted \cite{Cameron1993}, whereas PVLAS chose to rotate the magnetic field direction and uses now permanent dipole magnets, which are relatively inexpensive to buy, have no running costs and have in principle 100\% duty cycle, allowing very long integration times.

We are working with a Nd:YAG laser emitting radiation at $1064$~nm. Frequency doubled versions exist and could double the induced ellipticity, but at the moment the highest finesses have been obtained for infrared light.

The expected ellipticity must be extracted from the noise within the maximum available integration time. For this reason, the magnetic field is made time dependent, to move away from dc and limit $1/f$ noise. Homodyne and heterodyne detections are two possible detection schemes. This second technique has been adopted in the PVLAS apparatus.

\begin{figure}[htb]
\begin{center}
\includegraphics[width=10cm]{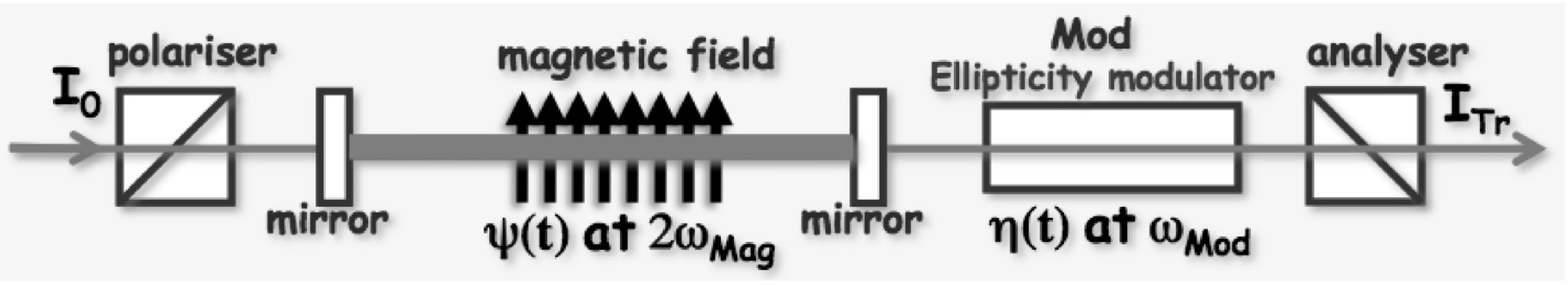}
\end{center}
\caption{Scheme of the PVLAS ellipsometer.}
\label{Scheme}
\end{figure}

A scheme of the PVLAS ellipsometer is shown in Figure \ref{Scheme}.
The input polariser linearly polarises the laser beam of intensity $I_{0}$ which then enters the sensitive region delimited by the Fabry-Perot cavity mirrors. The laser is phase locked to this cavity. After the cavity the laser beam passes through an ellipticity modulator which adds a known time dependent ellipticity $\eta(t)$ to the beam. This modulator ellipticity adds to the ellipticity $\Psi(t)$ acquired within the magnetic field region. After the modulator, the beam passes through the analyser which selects the polarisation perpendicular to the input polarisation and finally a photodiode detects $I_{\rm Tr}$. The sought for information can be extracted from the Fourier spectrum of $I_{\rm Tr}$ and from the value of the intensity $I_{\rm out}$ before the analyser.

\subsection{Estimate of the effect}

To better understand what follows, it is useful to present some numerical values of the different quantities involved in the PVLAS experiment. Considering the vacuum magnetic birefringence due to the Euler-Heisenberg-Weisskopf Lagrangian, let us determine the ellipticity we expect in the apparatus under construction. The magnets have a total magnetic field length $L = 1.6$~m with a measured field intensity $|\vec B_{\rm ext} | = 2.5$ T resulting in $\int{B^{2}_{\rm ext}}\,dL = 10$~T$^2$m. In the estimate we use the the maximum finesse value we have reached: ${\cal F} = 414000$. Putting these numbers together leads to
\begin{equation}
\Psi_{\rm PVLAS} = 2{\cal F}\,\frac{3 A_{e} \int{B^{2}_{\rm ext}}\,dL}{\lambda} = 3\times 10^{-11}.
\label{psipvlas}
\end{equation}
Assuming a maximum integration time $T_{\rm max} = 10^{6}$ s and a signal to noise ratio SNR = 1 implies that the sensitivity must be 
\begin{equation}
s_{\rm PVLAS} < \Psi_{\rm PVLAS}\sqrt{T_{\rm max}} = 3\times 10^{-8}~{\rm Hz}^{-1/2}.
\end{equation}
In principle this is well above the shot noise limit (see formula (\ref{shot}) below). Unfortunately, as will be briefly discussed below, other noise sources are present which limit the ellipsometer sensitivity. We now discuss in detail various issues of the measurement technique and of the setup.

\subsection{Heterodyne technique}

\begin{figure}[htb]
\begin{center}
\includegraphics[width=10cm]{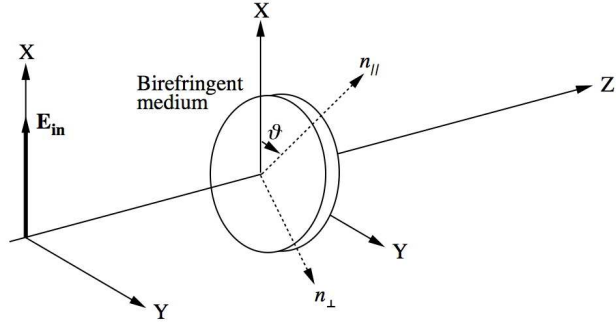}
\end{center}
\caption{Reference frame for the calculations using the Jones matrix formalism. The birefringent medium has a thickness $L$.}
\label{SistRif}
\end{figure}
Considering the coherence of the light source, a full treatment of the system can be done with the Jones matrix formalism \cite{Jones}. For the purpose of our discussion let the laser beam propagate along the $Z$ axis and let the incoming (linear) polarisation define the $X$ axis (Figure \ref{SistRif}). The Jones matrix for a uniaxial birefringent element of thickness $L$ is given by
\begin{equation}
\mathbf{BRF}(\vartheta)=
\left(
\begin{array}{cc}
1+\imath\psi\cos{2\vartheta} & \imath\psi\sin{2\vartheta} \\
 \imath\psi\sin{2\vartheta} & 1-\imath\psi\cos{2\vartheta}
 \end{array}
 \right)
 \end{equation}
where $\psi=\Delta\varphi/2$ ($\psi\ll1$) is now the maximum ellipticity acquired by the light and $\vartheta$ represents the angle between the slow axis ($n_\parallel > n_{\perp}$) of the medium and the $X$ axis. 

Let $\vec{E}_{\rm in}=E_{0}{1 \choose 0}$ be the entrance electric field; after the magnetic field region one has
\begin{eqnarray}
\nonumber
\vec{E}_{\rm BRF}&=&E_{0}\cdot\mathbf{BRF}\cdot{1 \choose 0} = E_{0}
\left(
\begin{array}{c}
1+\imath\psi\cos{2\vartheta} \\
\imath\psi\sin{2\vartheta}
\end{array}
\right).
\end{eqnarray}
Assuming no losses and adding the analyser --- a polariser crossed with respect to the entrance polariser, with Jones matrix $\mathbf{A}=\left(\begin{array}{cc}0&0\\0&1\end{array}\right)$ --- the intensity $I_{\rm Tr}$ will be
\begin{equation}
I_{\rm Tr}=I_{\rm 0}\left|\imath\psi\sin{2\vartheta}\right|^{2}.
\end{equation}
Given the predicted value, the resulting output power, proportional to $\psi^2$, is immeasurably small.

By introducing an ellipticity modulator (in our case a Photoelastic Modulator, PEM), which adds a known sinusoidal ellipticity $\eta(t)$, the detected signal is linearised in $\psi$. In fact, the Jones matrix for the modulator is the same as $\mathbf{BRF}$ with $\vartheta$ set at an angle of $\pi/4$ ($\psi\ll\eta\ll1$):
\begin{equation}
\mathbf{MOD}=
\left(
\begin{array}{cc}
1& \imath\eta(t) \\
 \imath\eta(t) & 1
 \end{array}
 \right).
 \end{equation}
The resulting vector describing the electric field after the modulator will be 
\begin{equation}
\vec{E}_{\rm MOD}=E_{\rm 0}\cdot\mathbf{MOD}\cdot\mathbf{BRF}\cdot{1 \choose 0}
\nonumber=E_{\rm 0}
\left(
\begin{array}{c}
1+\imath\psi\cos{2\vartheta}-\psi\eta(t)\sin2\vartheta \\
\imath\eta(t)+\imath\psi\sin{2\vartheta}-\eta(t)\psi\cos2\vartheta
\end{array}
\right).
\end{equation}
Neglecting second order terms, the intensity $I_{\rm Tr}$ after the analyser will be
\begin{equation}
I_{\rm Tr}(t)=I_{\rm 0}\left|\imath\eta(t)+\imath\psi\sin{2\vartheta}\right|^{2} \simeq I_{\rm 0}\left[\eta(t)^2+2\eta(t)\psi\sin2\vartheta\right]
\end{equation}
which now depends linearly on the ellipticity $\psi$. To complete the discussion, one finds experimentally that static and slowly varying ellipticities, indicated as $\alpha(t)$, are always present in a real apparatus and that two crossed polarisers have an intrinsic extinction ratio $\sigma^2$, mainly due to imperfections in the calcite crystals. Furthermore, losses in the system reduce the total light reaching the analyser, which we indicate as $I_{\rm out}$. Therefore, taking into account an additional spurious ellipticity term $\alpha(t)$ (since $\alpha, \psi, \eta \ll 1$ these terms commute and therefore add up algebraically) and a term proportional to $\sigma^2$, the total intensity at the output of the analyser will be
\begin{eqnarray}
\nonumber
I_{\rm Tr}(t)&=&I_{\rm out}\left[\sigma^2+\left|\imath\eta(t)+\imath\psi\sin{2\vartheta}+\imath\alpha(t)\right|^{2}\right]\simeq\\
&\simeq&I_{\rm out}\left[\sigma^2+\eta(t)^2+\alpha(t)^{2}+2\eta(t)\psi\sin2\vartheta+2\eta(t)\alpha(t)\right].
\end{eqnarray}

In PVLAS, to be able to distinguish the large term $\eta(t)\alpha(t)$ from the term $\eta(t)\psi\sin2\vartheta$, $\psi\sin2\vartheta$ is also modulated in time by rotating the magnetic field direction (varying $\vartheta$). The final expression, explicitly indicating the time dependence of $\vartheta$, for the intensity at the output of the analyser is therefore
\begin{equation}
I_{\rm Tr}(t)=I_{\rm out}\left[\sigma^2+\eta(t)^2+\alpha(t)^2+2\eta(t)\psi\sin2\vartheta(t)+2\eta(t)\alpha(t)\right].
\label{spurious}
\end{equation}

\subsection{Optical path multiplier}

To increase the ellipticity induced by the birefringent region of thickness $L$ one can increase the number of passes through it. Either a multi-pass cavity or a Fabry-Perot cavity can be used for this purpose. In the PVLAS experiment a Fabry-Perot has been chosen. In a multi-pass cavity the induced ellipticity is multiplied by the number of passes $N_{\rm pass}$ through the region. With a Fabry-Perot cavity the analogy to a multi-pass cavity is 
 not immediate since one is dealing with a standing wave.

Let $t$, $r$ be the transmission and reflection coefficients, and $p$ the losses of the mirrors of the cavity such that $t^2+r^2+p^2=1$. Let $d$ be the length of the cavity and $\delta={4\pi d}/{\lambda}$ the roundtrip phase for a beam of wavelength $\lambda$. Then the Jones matrix for the elements of the ellipsometer after the entrance polariser is
\begin{equation}
\mathbf{ELL}=
\mathbf{A}\cdot\mathbf{SP}\cdot\mathbf{MOD}\cdot{\it t}^{2}e^{\imath\delta/2}\sum_{n=0}^{\infty} {\left[\mathbf{BRF}^{2}{\it r}^{2}e^{\imath\delta}\right]^{n}}\cdot\mathbf{BRF}
 \end{equation} 
where $\mathbf{SP}$ describes spurious ellipticities. Because ${\it r}^{2} < 1$, $\mathbf{ELL}$ can be rewritten as
\begin{equation}
\mathbf{ELL}=
\mathbf{A}\cdot\mathbf{SP}\cdot\mathbf{MOD}\cdot{\it t}^{2}e^{\imath\delta/2}{\left[\mathbf{I}-\mathbf{BRF}^{2}{\it r}^{2}e^{\imath\delta}\right]^{-1}}\cdot\mathbf{BRF}
\label{fp}
\end{equation} 
with $\mathbf{I}$ the identity matrix. With the laser phase locked to the cavity so that $\delta=2\pi m$, where $m$ is an integer number, 
the electric field at the output of the system will be
\begin{equation}
\vec{E}_{\rm Tr}=E_{\rm 0}\cdot\mathbf{ELL}\cdot{1 \choose 0}
=E_{\rm 0}\frac{{\it t}^{2}}{{\it t}^{2}+p}
\left(
\begin{array}{c}
0 \\
\imath\alpha(t)+\imath\eta(t)+\imath\displaystyle\frac{1+{\it r}^{2}}{1-{\it r}^{2}}\psi\sin{2\vartheta}
\end{array}
\right)
\end{equation}
and the intensity, including mirror losses and polariser extinction,
\begin{equation}
I_{\rm Tr}(t)=I_{\rm out}\left[\sigma^2+\Big|\imath\alpha(t)+\imath\eta(t)+\imath\left(\frac{1+{\it r}^{2}}{1-{\it r}^{2}}\right)\psi\sin{2\vartheta}\Big|^{2}\right].
\label{Iout}
\end{equation}
This expression is at the basis of the ellipsometer in the PVLAS apparatus \cite{APB2006}.
Small ellipticities add up algebraically and the Fabry-Perot multiplies the single pass ellipticity $\psi\sin{2\vartheta}$, generated within the cavity, by a factor $({1+{\it r}^{2}})/({1-{\it r}^{2}})\approx{2{\cal F}}/{\pi}$, where ${\cal F}$ is the finesse of the cavity. The ellipticity signal to be detected is therefore $\Psi(\vartheta)=({2{\cal F}}/{\pi})\psi\sin{2\vartheta}$.

The Fabry-Perot as an optical path amplifier works well until the accumulated phase difference due to birefringences within the cavity is less than $\pi/2$. Beyond this value the interferometer has two separate resonances for two orthogonal polarisations \cite{APB2006} and therefore elliptically polarised light cannot be resonant. As total birefringence also includes the intrinsic birefringence of the cavity mirrors, two mirrors with similar characteristics must be used and their birefringence axes aligned.

\subsection{Fourier components}

In the PVLAS experiment, $\eta(t)=\eta_{0}\cos(2\pi\nu_{\rm Mod}t+\theta_{\rm Mod})$ and the magnetic field direction is rotated at a frequency $\nu_{\rm Mag}$: $\vartheta(t)=2\pi\nu_{\rm Mag}t+\vartheta_{\rm Mag}$. A Fourier analysis of the intensity $I_{\rm Tr}(t)$ of equation (\ref{Iout}) results in four main frequency components each with a definite amplitude and phase. These are reported in table \ref{components}.
\begin{table}[h!]
\caption{Intensity of the frequency components of the signal after the analyser.}
\begin{center}
\begin{tabular}{c|c|c|c}
\hline
Frequency & Fourier component & Intensity/$I_{\rm out}$ & Phase\\
\hline
dc & $I_{\rm dc}$ & $\sigma^2+\alpha_{\rm dc}^2+\eta_{0}^{2}/2$ & $-$\\
$\nu_{\rm Mod}$ & $I_{\nu_{\rm Mod}}$ & $2\alpha_{\rm dc}\eta_{0}$ & $\theta_{\rm Mod}$\\
$\nu_{\rm Mod}\pm2\nu_{\rm Mag}$ & $I_{\nu_{\rm Mod}\pm2\nu_{\rm Mag}}$ & $\eta_{0}\displaystyle\frac{2{\cal F}}{\pi}\psi$ & $\theta_{\rm Mod}\pm2\vartheta_{\rm Mag}$\\
$2\nu_{\rm Mod}$ & $I_{2\nu_{\rm Mod}}$ & $\eta_{0}^{2}/2$ & $2\theta_{\rm Mod}$\\
\hline
\end{tabular}
\end{center}
\label{components}
\end{table}

The presence of a component at $\nu_{\rm Mod}\pm2\nu_{\rm Mag}$ in the signal identifies an ellipticity induced by the rotating magnetic field. Furthermore the phase of this component must satisfy the value in table \ref{components}. In the presence of a signal above background with the correct Fourier phase, the ellipticity $\Psi=(2{\cal F}/\pi)\psi$ can be calculated from $I_{\rm out}$, from the Fourier components
$I_{\nu_{\rm Mod}\pm2\nu_{\rm Mag}}$, and from $I_{2\nu_{\rm Mod}}$ as the average of the two
sideband signals:
\begin{equation}
\Psi=\frac{1}{2}\left(\frac{I_{\nu_{\rm Mod}+2\nu_{\rm Mag}}}{\sqrt{2I_{\rm out}I_{2\nu_{\rm Mod}}}}+\frac{I_{\nu_{\rm Mod}-2\nu_{\rm Mag}}}{\sqrt{2I_{\rm out}I_{2\nu_{\rm Mod}}}}\right).
\end{equation}

\subsection{Noise considerations}

Indicating with $R_{\nu_{\rm Mod}\pm2\nu_{\rm Mag}}$ the noise spectral density at the signal frequencies, and assuming $R_{\nu_{\rm Mod}+2\nu_{\rm Mag}}=R_{\nu_{\rm Mod}-2\nu_{\rm Mag}}$, the sensitivity spectral density $s$ of the ellipsometer for a unity signal to noise ratio is
\begin{equation}
s=\frac{R_{\nu_{\rm Mod}+2\nu_{\rm Mag}}}{\sqrt{4I_{\rm out}I_{2\nu_{\rm Mod}}}}.
\label{sensitivity}
\end{equation}

In principle, the noise limit for such a system is determined by the rms shot noise $i_{\rm shot}$ of the current $i_{\rm dc}$:
\begin{equation}
i_{\rm shot}=\sqrt{2ei_{\rm dc}\Delta\nu}=\sqrt{2eI_{\rm out}q\left(\sigma^2+\frac{\eta_0^2}{2}+\alpha_{\rm dc}^2\right)\Delta\nu}
\end{equation}
where $q$ is the efficiency of the photodetector, and $\Delta\nu$ the bandwidth. In the case $\eta_0^2\gg\sigma^2,\alpha_{\rm dc}^2$, the dc current will depend only on $\eta_0$ and by substituting $R_{\nu_{\rm Mod}+2\nu_{\rm Mag}}=i_{\rm shot}/(q\sqrt{\Delta\nu})$ into Eq. (\ref{sensitivity}), the shot-noise sensitivity spectral density $s_{\rm shot}$ becomes
\begin{equation}
s_{\rm shot}=\sqrt{\frac{e}{I_{\rm out}q}}.
\label{shot}
\end{equation}
Assuming an intensity $I_{\rm out} = 5$~mW and the efficiency of the diode $q = 0.7$~A/W, equation above gives $s_{\rm shot}\sim 7\times10^{-9}$~Hz$^{-1/2}$.

Other intrinsic noise sources are photodiode dark current noise $i_{\rm dark}$ which generates an output noise signal density $V_{\rm dark}=G\,i_{\rm dark}$, with $G$ the transimpedance gain of the photodetection system, Johnson rms current noise spectral density in the transimpedance resistor in the amplifier of the photodiode $i_{\rm J}=\sqrt{4k_BT/G}$, and relative laser intensity current noise spectral density $i_{\rm RIN}=I_{\rm out}\,q\cdot{\rm RIN}(\nu)$, where ${\rm RIN}(\nu)=I_{\rm noise}(\nu)/I_0$ is the relative amplitude noise spectral density of the laser light. These noises must be kept below $i_{\rm shot}$ at a frequency near $\nu_{\rm Mod}$ in order to reach the theoretical sensitivity. Complete expressions for these noise contributions to the ellipticity spectral noise density can be obtained from Eq. (\ref{sensitivity}) as functions of the modulation amplitude $\eta_0$:
\begin{eqnarray}
s_{\rm shot}&=&\sqrt{\frac{2e}{I_{\rm out}q}\left(\frac{\sigma^2+\eta_0^2/2}{\eta_0^2}\right)}\\
s_{\rm dark}&=&\frac{V_{\rm dark}}{G}\,\frac{1}{I_{\rm out}\,q\,\eta_0}\\
s_{\rm J}&=&\sqrt{\frac{4k_BT}{G}}\,\frac{1}{I_{\rm out}\,q\,\eta_0}\\
s_{\rm RIN}&=&{\rm RIN}(\nu_{\rm Mod})\,\frac{\sqrt{(\sigma^2+\eta_0^2/2)^2+(\eta_0^2/2)^2}}{\eta_0}.
\end{eqnarray}

\section{Experimental studies}

\subsection{PVLAS - LNL: main features and limitations}

The previous PVLAS apparatus was set up at LNL in Legnaro, Italy and took data from 2002 to 2007. It featured a vertical assembly of the ellipsometer with a 6~m long Fabry-Perot cavity, the injection and detection optics installed on different optical benches resting on the ground and on the top of a granite tower nearly 8~m high, respectively. A superconductive dipole magnet about 1~m long, installed in a cryostat resting on a turntable, provided a rotating magnetic field orthogonal to the light path in the Fabry-Perot cavity. The magnet was operated with magnetic fields up to 5.5~T.
The magnet support and the optics tower had different foundations, and were hence mechanically decoupled.
Although this apparatus has set best limits on magnetic vacuum birefringence and photon-photon elastic scattering at low energies \cite{Bregant2008}, it suffered from several limitations:

\begin{itemize}
\item high stray field when operating the superconducting magnet above 2.3~T, due to saturation of the iron return yoke;
\item limited running time due to liquid Helium consumption;
\item high running costs for liquid Helium;
\item the granite tower could not be seismically insulated, due to its size and configuration;
\item with a single magnet, a zero ellipticity measurement was only possible with field off, hence not in the same experimental condition as in the magnetic birefringence measurements.
\end{itemize}

To solve the first three problems, we have now chosen to work with permanent magnets instead of superconducting ones. In permanent magnets the stray field can be made much smaller and the duty cycle can be as high as 100\%. Seismic noise also limited the sensitivity of the LNL setup; instead, a large improvement of this parameter can be obtained by installing the whole ellipsometer on a single seismically isolated optical bench, as shown in ref. \cite{towardsQED}. The last point in the list above is extensively addressed in a separate section.

A completely new setup has been designed according to these lines and is being installed in a clean room of the Ferrara section of INFN. The new set up is built using a single 4.8~m long granite optical bench actively isolated, two identical rotating permanent dipole magnets in Halbach configuration, each 0.8~m in length with magnetic field strength $B = 2.5$~T and 2~cm bore. The magnets design features a magnetic field shield strongly reducing stray fields. We plan to rotate them at $\nu_{\rm Mag}=10$~Hz, thanks to their reduced mass and good balancing. The optical setup features a 2~W tuneable Nd:YAG laser frequency-locked to a 3.2~m long Fabry-Perot cavity using a modified Pound-Drever-Hall scheme \cite{Cantatore1995} and an entirely non magnetic optomechanical setup. The motion of the in-vacuum optical elements is based on stepper piezo motors which maintain their position when powered off. The whole ellipsometer is kept in a vacuum chamber pumped with non evaporable getters, which are intrinsically non magnetic. Two 1~m long glass tubes traverse the bore of the magnets; each tube is pumped at both ends.

\subsection{Two magnet configuration}

As reported in ref. \cite{PRD2008}, the rotating stray magnetic field and the vibrations associated to the masses in rotation can act on the optics and fake ellipticity signals. It is therefore desirable to perform measurements of the zero of the ellipticity scale in conditions as close as possible to those in which an authentic signal is expected. This is not possible with one single superconducting magnet, because when there is no field in the magnet bore, also the stray field is absent. By using two identical magnets and orienting their fields at $90^{\circ}$ the net ellipticity generated by the magnetic birefringence is zero. Running the system with the magnets parallel and perpendicular will allow the identification of a real physical signal with respect to some spurious signal due to stray field.

\begin{figure}[htb]
\begin{center}
\includegraphics[width=10cm]{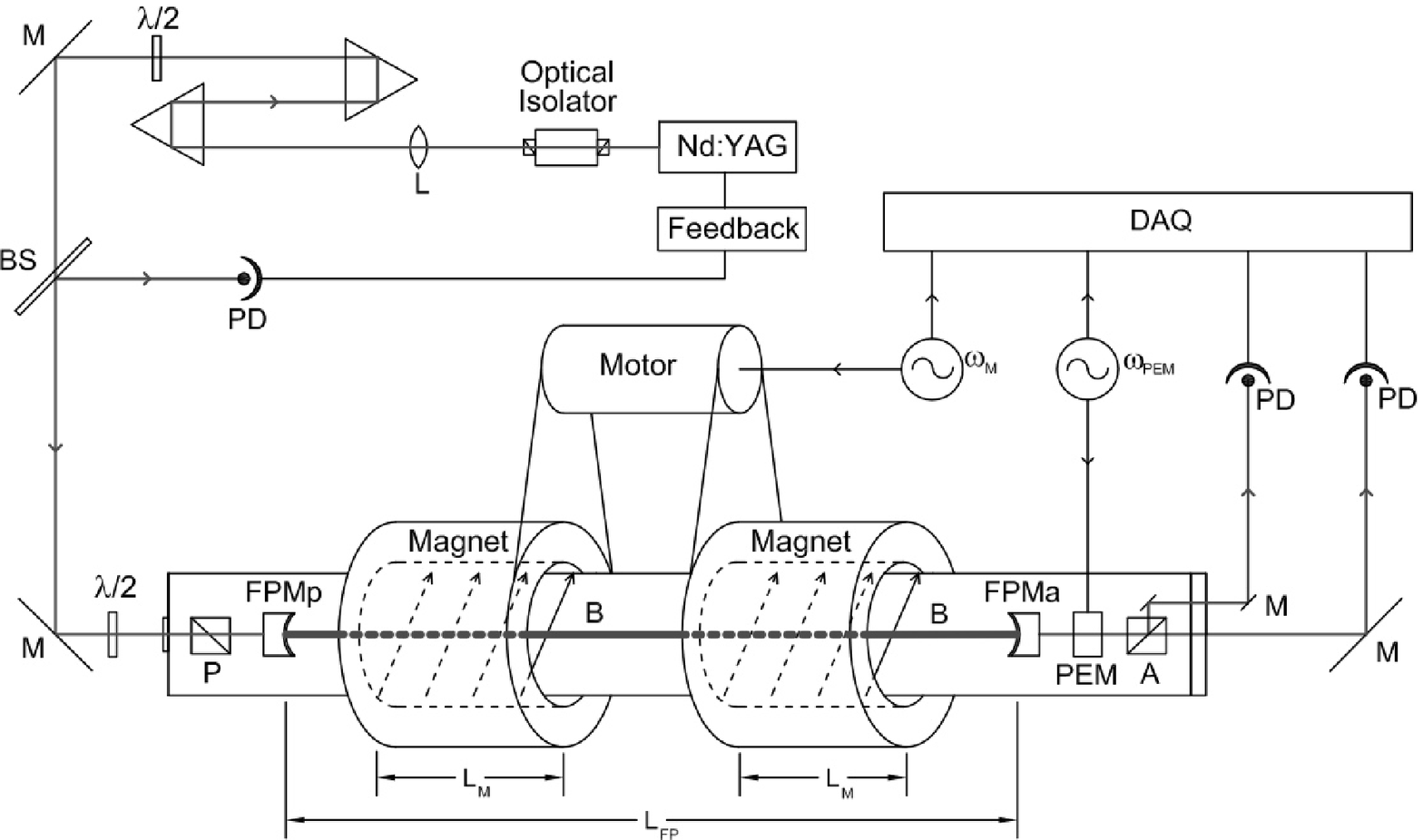}\\
\includegraphics[width=10cm]{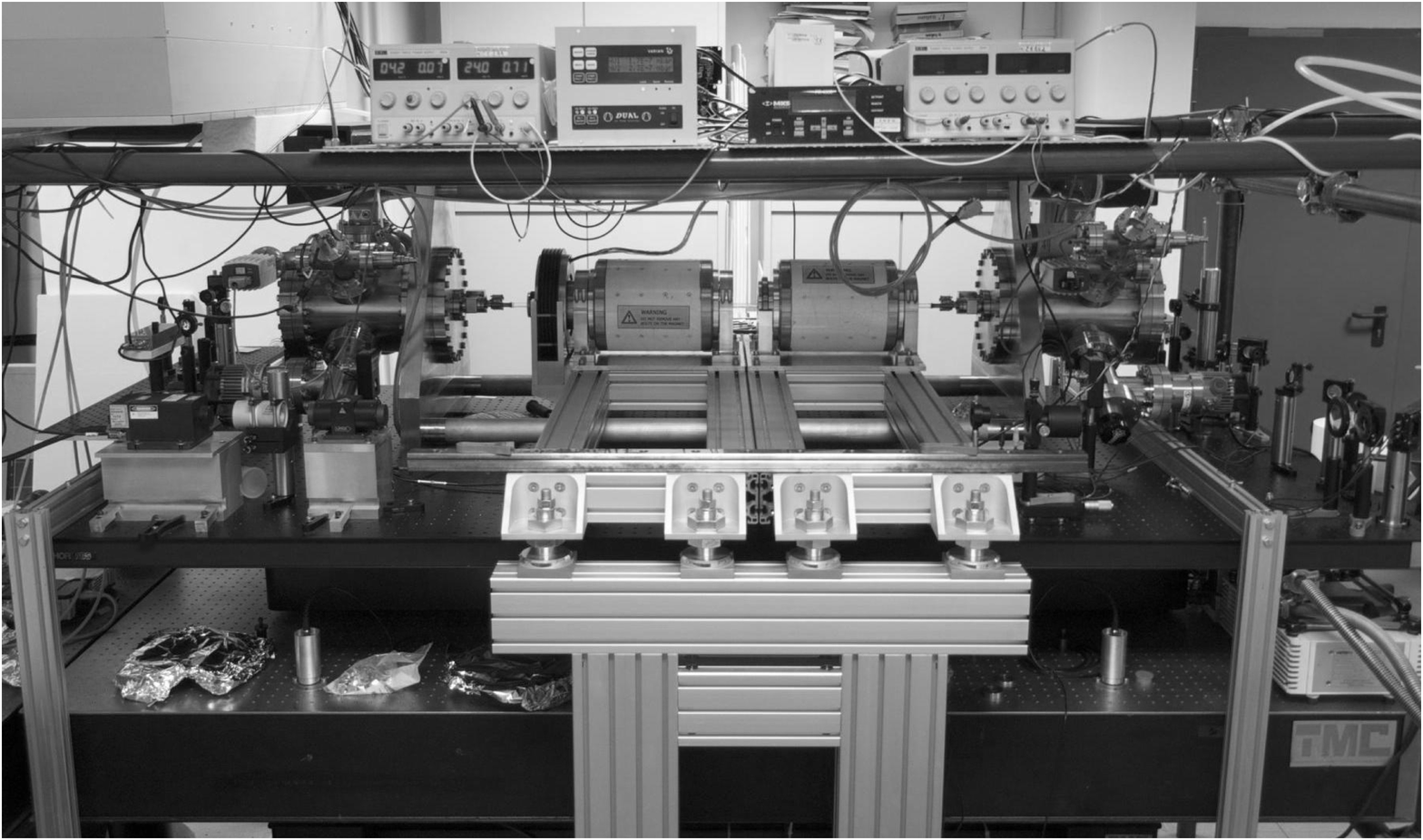}
\end{center}
\caption{Optical scheme (upper panel) and photograph (lower panel) of the test apparatus in Ferrara. At the centre one can see the two dipole permanent magnets. The optics is supported by antivibration stages whereas the magnet supports are on the floor.}
\label{testsetup}
\end{figure}
This idea has been verified with the scaled down test setup using the Cotton-Mouton effect of oxygen \cite{Brandi1998}. The principle of the test ellipsometer is identical to that of the final system in construction. Fig. \ref{testsetup}a shows the scheme of the test apparatus and Fig. \ref{testsetup}b a picture of it. At the centre one can see the two permanent magnets each generating a 20~cm long magnetic field of maximum intensity $B = 2.3 $~T. For these magnets the measured $\int{B^{2}\,dL}$ is 1.85~T$^2$m.
The whole optical setup is placed on a seismically isolated optical bench, whereas the magnets are supported by a separate structure resting on the floor and thereby mechanically isolated from the optics.
The finesse of the cavity was ${\cal F} = 240000$ and the oxygen pressure inside the apparatus was 0.278 mbar.

\begin{figure}[htb]
\begin{center}
\includegraphics[width=10cm]{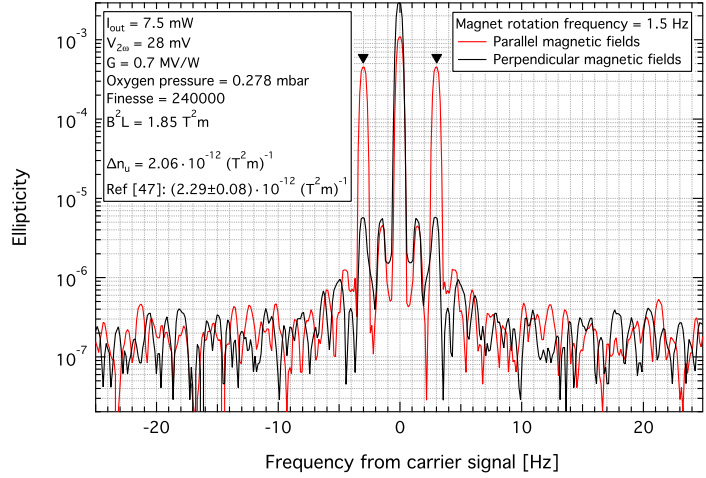}
\end{center}
\caption{Fourier spectrum around the carrier frequency $\nu_{\rm Mod}$. Ellipticity measurements with the magnets parallel (red) and with the magnets perpendicular (black). The large sidebands at $2\nu_{\rm Mag}$ are due to the Cotton Mouton effect in oxygen gas. No error is given on the measured value of $\Delta n_u$ (for a definition see Ref. \cite{Rizzo1997}), as we could not control gas composition.}
\label{CM_O2}
\end{figure}
Figure \ref{CM_O2} shows the ellipticity Fourier spectrum around the carrier frequency $\nu_{\rm Mod}$. Clear sidebands can be seen at twice the rotation frequency of the magnets ($\nu_{\rm Mag} = 1.5$~Hz). Shown in red is the spectrum with the magnets in a parallel configuration whereas the black curve refers to the magnets perpendicular. The signal attenuation factor in the perpendicular configuration with respect to the parallel one is about 80: we can conclude that the parameter $B_{\rm ext}^{2}L$ for the two magnets is equal to within about 1-2\% and, most important, that the principle is correct. The presence of the unwanted sidebands at $\nu_{\rm Mag}$ indicates that a small spurious component could also contribute to the $2\nu_{\rm Mag}$ peak, thus limiting the cancellation of the signal in the perpendicular configuration.

We believe the improvement of using two magnets instead of one will be crucial in understanding the ellipsometer, and that such a measurement with two rotating magnets whose relative field directions can be changed is imperative. This is a straightforward way of guaranteeing the authenticity of a signal.

\subsection{Vacuum ellipticity measurements with the test setup}

With the test setup (Figure \ref{testsetup}), featuring two small magnets, measurements have been performed to understand its limits and optimise the new apparatus in construction.
Two different noise sources exist and are under study: wide-band noise and spurious signals at the magnet rotation frequency and its harmonics. Below we briefly report some results in the two cases.

\subsubsection{Sensitivity - wide-band noise}

Noise measurements were first performed without the Fabry-Perot cavity. In this configuration we successfully excluded noise sources from readout electronics and optical elements other than the cavity mirrors, practically reaching the expected sensitivity of $s_{\rm no~cavity} = 6\times10^{-9}\;{\rm Hz}^{-1/2}$, dominated by $s_{\rm RIN}$, and a noise floor of $\Psi_{\rm floor} = 1.5\times10^{-10}$ with 1600~s integration time.

\begin{figure}[htb]
\begin{center}
\includegraphics[width=10cm]{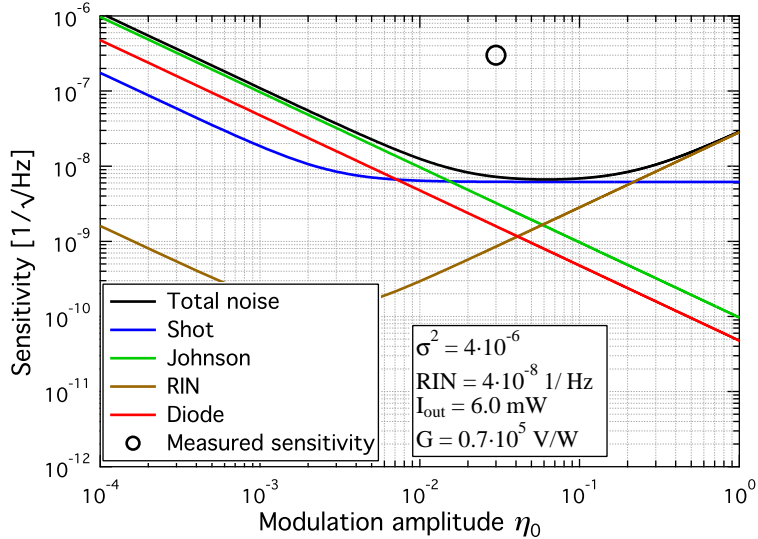}
\end{center}
\caption{Calculated noise contributions for the PVLAS test apparatus (solid curves). The green curve is the Johnson noise contribution from the transimpedance resistor in the current to voltage converter, the red curve is the dark current noise contribution from the photodiode, the blue curve is the shot noise contribution and the brown curve is the relative intensity noise contribution measured at the output of the cavity. The circle represents the measured ellipticity sensitivity of the apparatus.}
\label{noise}
\end{figure}
With the introduction of a cavity with finesse ${\cal F} = 240000$ the noise increased to $s_{\rm cavity} = 3\times 10^{-7}\;{\rm Hz}^{-1/2}$ at about 6~Hz. This was significantly more than what was expected from the reduction of $I_{\rm out}$ due to cavity losses (see Eq. (\ref{shot})). The actual ellipticity noise budget situation is depicted in Figure \ref{noise} as a function of the modulation amplitude $\eta_0$ \cite{Bregant2008}; in the figure, Relative Intensity Noise (RIN), Johnson noise in the amplifier transimpedance, intrinsic current noise in the photodiode and shot noise are summed quadratically. The circle marks the observed noise level. This unexplained noise is under study. We suspect variations of the intrinsic birefringence of the mirrors.

We also observed that the magnet rotation did not contribute to the wide-band noise, indicating a good isolation between the magnet support and the optical setup.

\subsubsection{Spurious peaks}

With the magnets in rotation we often observe ellipticity peaks varying from a few $10^{-8}$ to a few $10^{-7}$ whereas sometimes such peaks are not present. The frequencies of these peaks are at harmonics of the magnet rotation frequency. The variability of these peaks from one run to another seems to depend (in a non reproducible way) on the adjustment of the input and output polarisers which is done with motorised stages. To study the dependence of such peaks on the magnet orientation, a field probe is present near the output side magnet. Changing the relative orientation of the two magnets does not change the amplitude of these peaks but does change their phase; the result of this study indicates that the more sensitive part of the apparatus is the entrance optics. All the motorised stages have small electric motors which might couple to the rotating stray magnetic field and may introduce beam jitter and therefore ellipticity. The substitution of all these stages is under way.

\subsubsection{Noise floor measurements}

\begin{figure}[htb]
\hspace{0.39cm}
\begin{center}
\includegraphics[width=10.52cm]{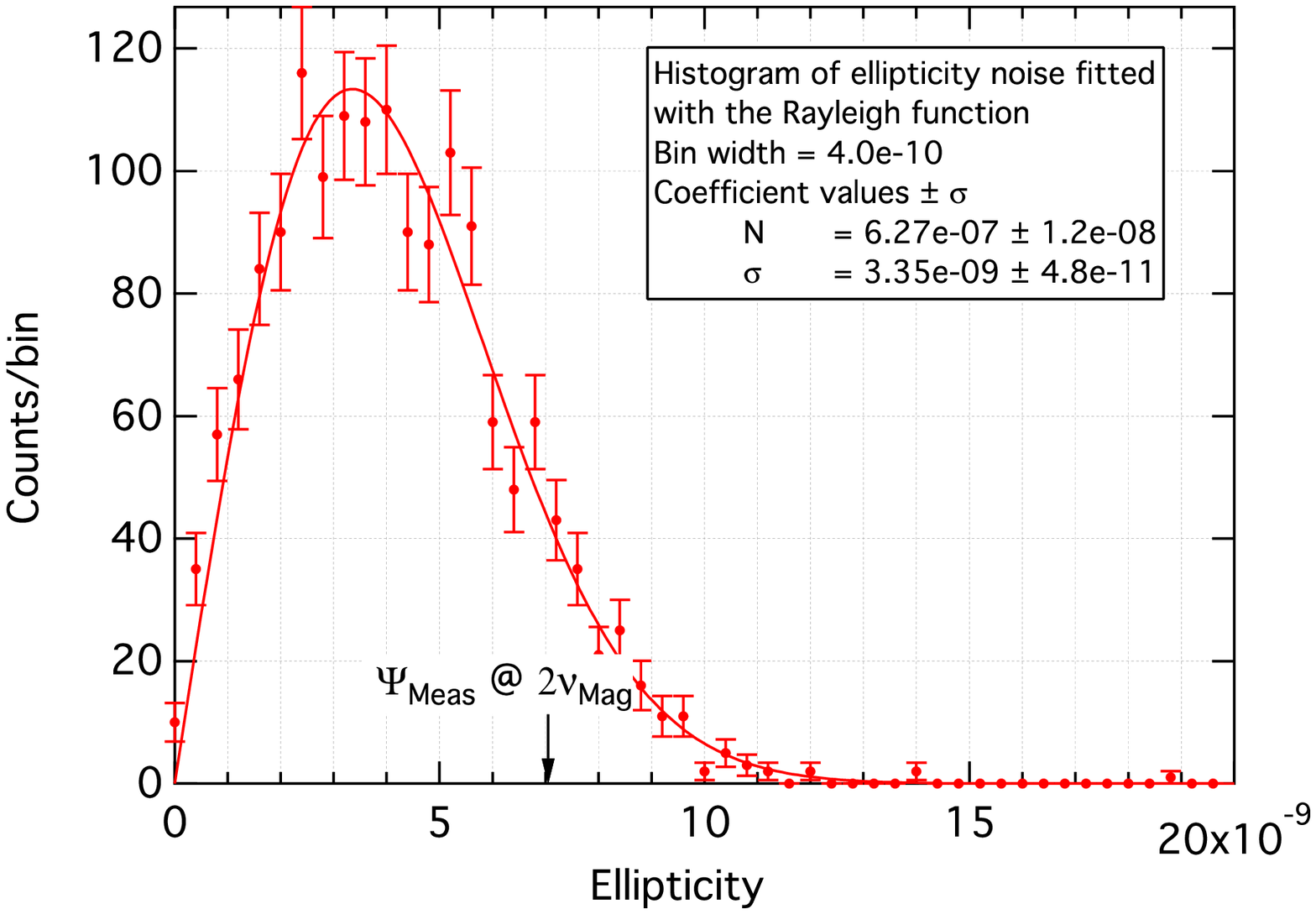}\\
\includegraphics[width=10cm]{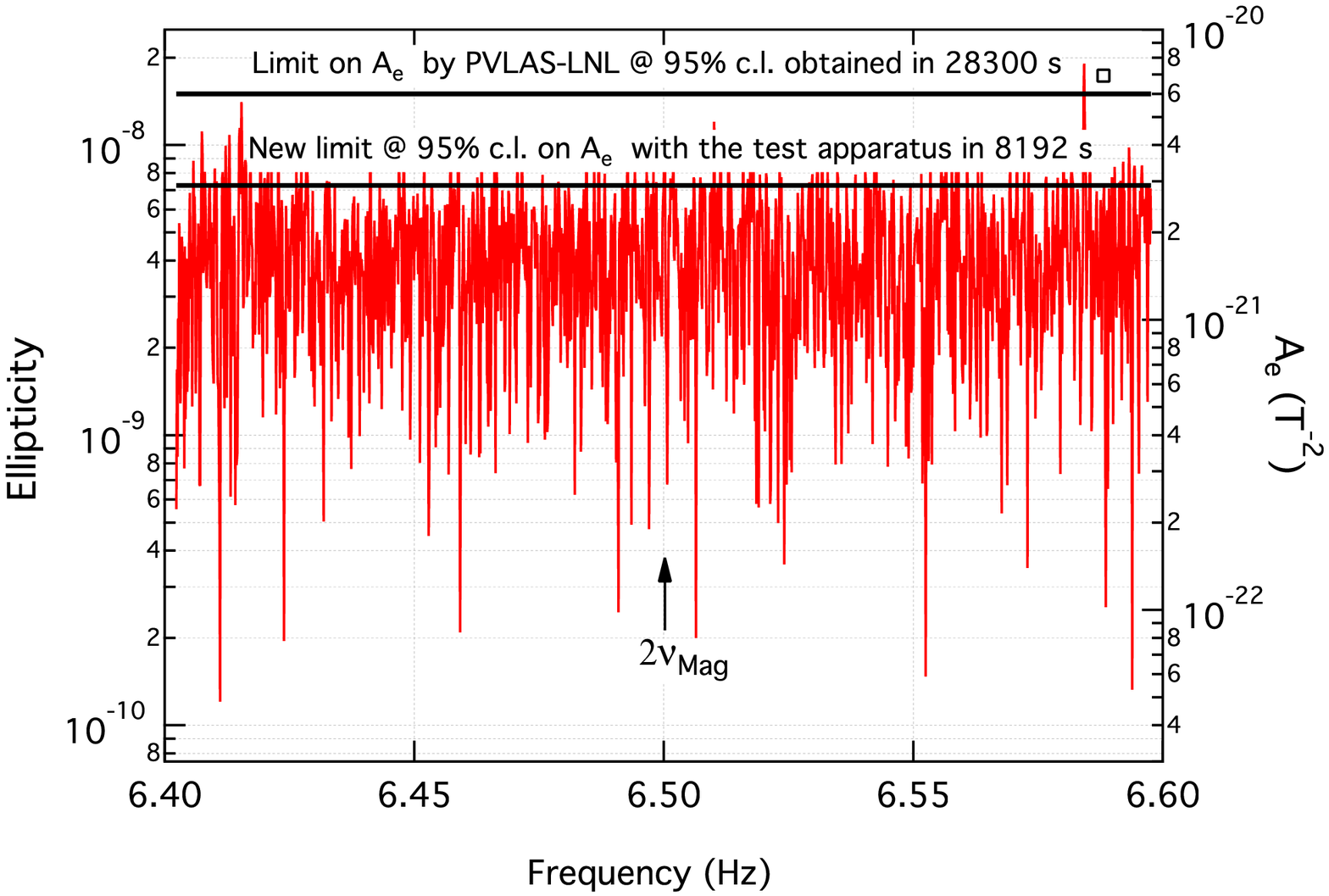}
\end{center}
\caption{Top: ellipticity noise spectrum in a 0.78~Hz frequency band around $2\nu_{\rm Mag}$. The spectrum is demodulated with a lock-in amplifier with respect to the $\nu_{\rm Mod}$ carrier. The integration time was $T = 8192$~s. Bottom: histogram of the ellipticity noise values plotted above. A fit with the Rayleigh distribution is superimposed with an ellipticity standard deviation $\sigma = 3.35\times 10^{-9}$. The arrow indicates the value of the ellipticity Fourier spectrum at exactly $2\nu_{\rm Mag} = 6.5$~Hz.}
\label{limite}
\end{figure}
With the test apparatus in a condition in which the peak at $2\nu_{\rm Mag}$ was not present, measurements of a few hours have been done. The magnet rotation frequency was $\nu_{\rm Mag} = 3.25$~Hz. In the top panel of Figure \ref{limite} we report the amplitude Fourier noise spectrum of the ellipticity in a frequency band of 0.78~Hz centred at $2\nu_{\rm Mag}$ for the best measurement; phase information is not reported.  The total integration time was 8192~s acquired in a single time record.

In the bottom panel of the figure, a 50 bins histogram of the same data is given. A vertical arrow in the same figure indicates the ellipticity value in the Fourier spectrum bin at exactly $2 \nu_{\rm Mag} = 6.5$~Hz. The probability density function for a noise signal with equal standard deviations $\sigma$ for the `in phase' and quadrature components is the Rayleigh function: $P(r) = N \frac{r}{\sigma^{2}}e^{-\frac{r^2}{2\sigma^2}}$. A fit with the Rayleigh function is superimposed to the histogram. The resulting ellipticity standard deviation is $\sigma = 3.35\times10^{-9}$ which translates, at 95\% c.l., in a birefringence limit induced by the magnetic field of
\begin{equation}
\Delta n < \sqrt{-2\ln(1-0.95)}\,\frac{\sigma\lambda}{2{\cal F}L} = 4.6\times10^{-20}.
\end{equation}
The limits obtained for the parameter $A_e$ and the photon-photon total elastic scattering cross section (at 95 \% c.l.) are:
\begin{eqnarray}
A_e& <& \frac{\Delta n}{3 \langle B_{\rm ext}^{2}\rangle} = 2.45\,\frac{\sigma\lambda}{6{\cal F}\int{B_{\rm ext}^{2}\,dL}} = 3.3\times10^{-21}\;{\rm T}^{-2}\\
\sigma_{\gamma\gamma}& <& 1.2\times10^{-62}\;{\rm m}^{2}\; @ \;1064\;{\rm nm}
\end{eqnarray}

\begin{figure}[htb]
\begin{center}
\includegraphics[width=10cm]{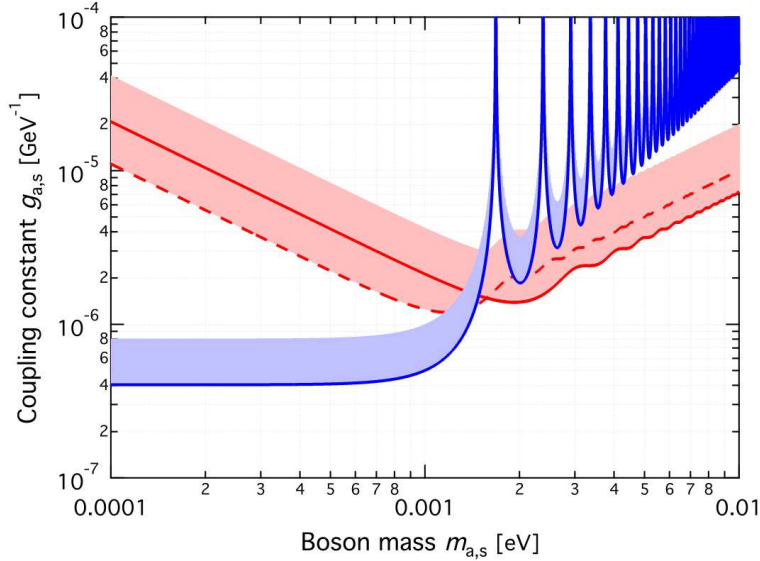}
\end{center}
\caption{PVLAS existence bounds for ALP's. The excluded region is above the curves. For $m_a\lesssim1.5$~meV the best limit is still given by the 2008 rotation measurement \cite{PRD2008}. Above this mass value a new limit is set by the measurement presented in this work. For the sake of comparison, the outdated 2008 ellipticity cutves is drawn (dashed red).}
\label{limitialp}
\end{figure}
\begin{figure}[htb]
\begin{center}
\includegraphics[width=9cm]{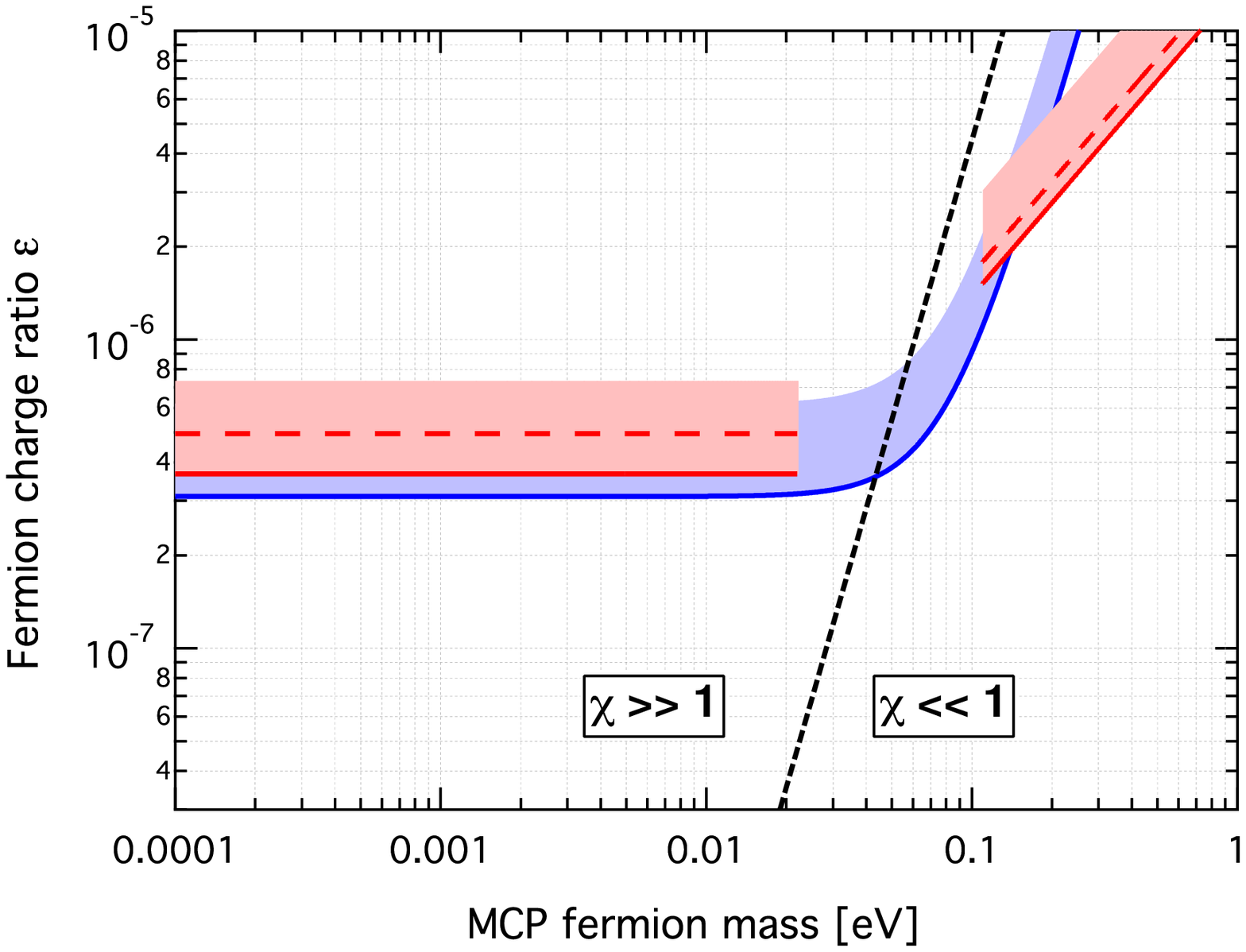}\\
\includegraphics[width=9cm]{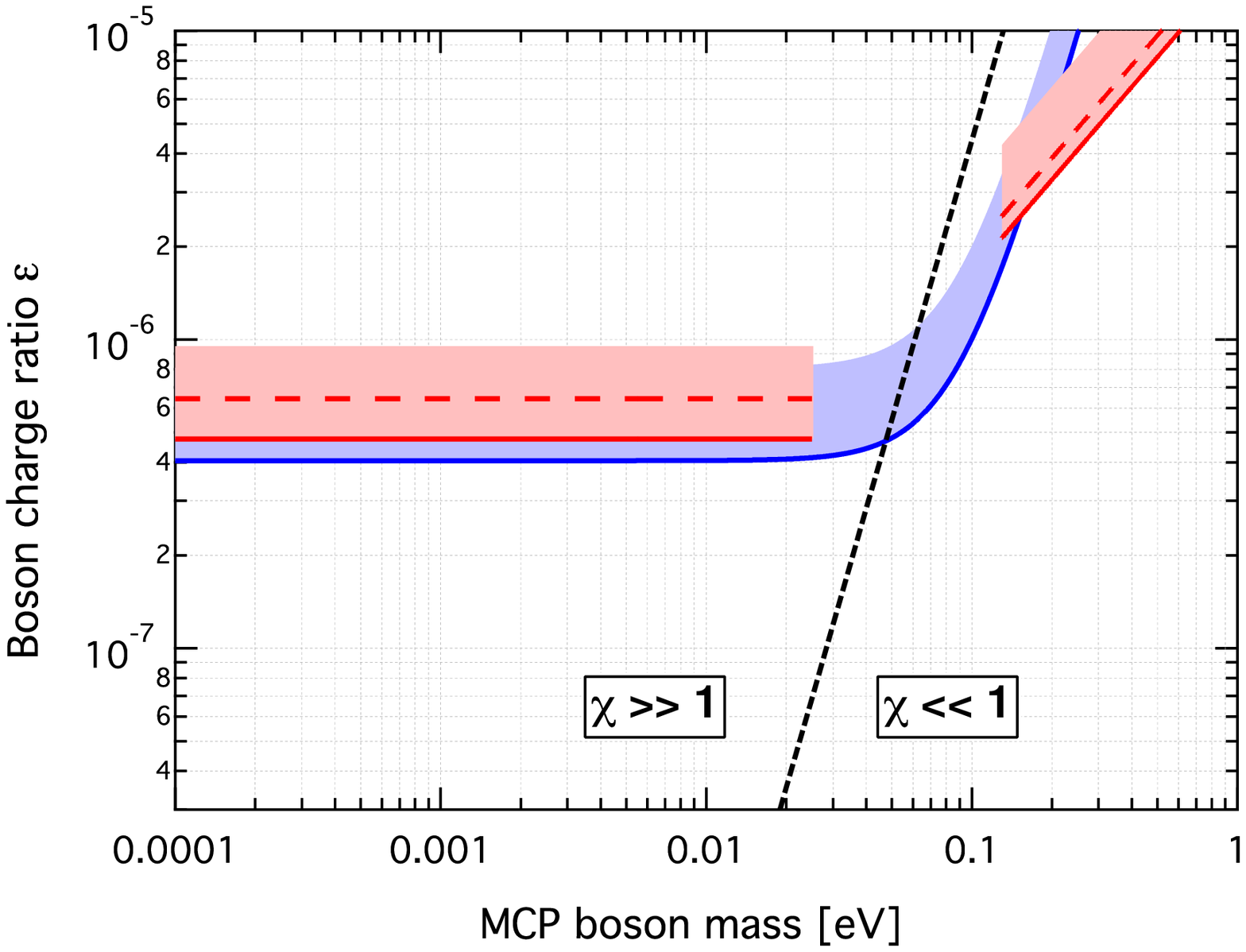}
\end{center}
\caption{PVLAS existence bounds for MCP's. The excluded regions are above the curves. Top: Dirac fermions; bottom: scalar bosons. In both cases the best limits are still set by the 2008 rotation measurement. The continuous blue rotation curves are taken from Ref. \cite{Ahlers2008}. The four ellipticity curves, drawn according to Eqs. (\ref{deltandf}) and (\ref{deltansc}), correspond to the outdated 2008 limit (dashed red) and to the new ellipticity limit (continuous red). The two branches of each of the ellipticity curves are not connected in the mass range around $\chi=1$ (dotted black line), where the difference of the indices of refraction changes sign.}
\label{limitimcp}
\end{figure}
As shown in Figure \ref{limitialp}, the ellipticity measurement presented here also sets new laboratory limits on the existence of axion-like particles for masses above 1.5 meV. Below 1.5 meV the best laboratory limit is still given by the PVLAS - LNL 2008 dichroism measurement \cite{PRD2008}. In the case of millicharged particles we confirm the best exclusion limits set by the 2008 rotation measurements (see Figs. \ref{limitimcp}) for both fermions and scalars.

\section{Other experimental efforts}

Three other experiments aimed at studying vacuum magnetic birefringence exist: the Q \& A experiment, the BMV experiment and the OSQAR experiment. All of them study the propagation of a laser beam through a magnetic field region and use a Fabry-Perot interferometer to increase the optical path length. A number of other proposals have been put forward to measure vacuum magnetic birefringence \cite{Hall2000,Luiten2004,Zavattini2009} or other related quantities \cite{Partovi1993} with different techniques: they will not be described here. As far as the OSQAR experiment \cite{Pugnat2005} is concerned, the information available in the literature at the time this article was written was rather incomplete.

As shown before, vacuum magnetic birefringence appears in a number of different theories. In the case of the Euler-Heisenberg-Weisskopf Lagrangian, $\Delta n$ is proportional to $B_{\rm ext}^2$ through the parameter $A_{e} = \Delta n/(3 B_{\rm ext}^{2})$. On the other hand, the optical quantity the experiments are measuring is the ellipticity generated by the induced birefringence. The relation between birefringence and ellipticity is given in formula (\ref{Psi}). The three experiments have therefore two distinct parts: the optical ellipsometer and the magnetic field source. Figures of merit for the optical setup are the ellipticity sensitivity $s$ at the frequency of interest, the signal amplification factor defined by the Fabry-Perot cavity finesse and the wavelength used.

\begin{table}[htb]
\begin{center}
\caption{Main parameters of the three ongoing experiments to measure vacuum magnetic birefringence. Effect modulation frequency $f_{\rm mod}$ is $2\nu_{\rm Mag}$ for PVLAS and Q~\&~A, and $1/T_{\rm pulse}$ for BMV.}
{\scriptsize
\begin{center}
\begin{tabular}{lccc}
\hline\hline
Experiment & PVLAS & Q \& A & BMV \\
\hline
Status & Achieved/Planned & Achieved/Planned & Achieved/Planned \\
\hline
Wavelength (nm)                 & 1064        & 1064/532   & 1064
\\
Magnetic dipole                 & permanent   & permanent  & pulsed
\\
$\int B^2\,dL$ (T$^2$m)          & 1.85/10     & 3.2/19     & 25/600
\\
Average $B_{\rm ext}$ (T)       & 2.15/2.5    & 2.3/2.3    & 14/30
\\
Finesse                         & $2.4\times10^5$/$ > 4\times10^5$      & $3\times10^4$/$1\times10^5$         & $5\times10^5$/$1\times10^6$
\\
QED ellipticity (Eq. \ref{Psi}) & $3\times10^{-12}$/$3\times10^{-11}$ & $7\times10^{-13}$/$3\times10^{-11}$ & $9\times10^{-11}$/$5\times10^{-9}$
\\
Detection scheme                & heterodyne  & heterodyne & homodyne
\\
Effect mod. freq. $f_{\rm mod}$     & 6 Hz/20 Hz & 26 Hz                  & 500 Hz
\\
Duty cicle $D_{t}$              & $\sim1$     & $\sim1$    & $3\times10^{-6}$
\\
$s$ @ $f_{\rm mod}~({\rm Hz}^{-1/2})$    & $3\times10^{-7}$/$3\times10^{-8}$                   & $1\times10^{-6}$/$1\times10^{-8}$   &  $5\times10^{-8}$/$7\times10^{-9}$
\\
$s_{\rm eff}$ @ $f_{\rm mod}~({\rm Hz}^{-1/2})$ & $3\times10^{-7}$/$3\times10^{-8}$   & $1\times10^{-6}$/$1\times10^{-8}$   &  $3\times10^{-5}$/$4\times10^{-6}$
\\
$\Delta n_{\rm eff}$ sensitivity $({\rm Hz}^{-1/2})$ & $1.7\times10^{-18}$/$2.5\times10^{-20}$ & $3.0\times10^{-17}$/$7.4\times10^{-21}$ & $2.6\times10^{-16}$/$3.0\times10^{-18}$
\\
$A_{\rm e}$ sensitivity $({\rm T}^{-2}{\rm Hz}^{-1/2})$ & $1.2\times10^{-19}$/$1.3\times10^{-21}$            & $1.8\times10^{-18}$/$4.7\times10^{-22}$ & $4.4\times10^{-19}$/$1.1\times10^{-21}$
\\
Time for SNR = 1                & 260 yr/12 d & $63\times10^3$ yr/1.4 d    & $3.6\times10^3$ yr/8.3 d
\\
\hline\hline
\end{tabular}
\end{center}}
\end{center}
\label{summary}
\end{table}

In Table \ref{summary} the main features of the ongoing experiments are summarised with an estimated running time necessary to reach this goal. As the scheme of the Q \& A experiment \cite{Ni1996} closely resembles PVLAS's one, the interpretation of its parameters is straightforward. The BMV experiment deserves a few more words: it employs an intense pulsed magnetic field lasting $T_{\rm pulse}=2$~ms FWHM (total duration about 4~ms), with a repetition rate of five shots per hour to allow the magnets to cool down. Therefore on average the effective duty cycle with the magnetic field ON is $D_{t}=2~{\rm ms}\times(5/3600~{\rm s})\approx3\times 10^{-6}$. The analysis correlates the measured ellipticity during each pulse, assumed to be proportional to $B_{\rm ext}^{2}$, with the magnetic pulse shape, also taking into account the dynamical response of the Fabry-Perot whose decay time is of the order of the magnetic pulse rise time \cite{Berceau2012}. The analysis gives $\Delta n_{\rm B} = \Delta n/B_{\rm filtered}^2$ for each pulse, where $B_{\rm filtered}$ is the effective field filtered by the first order response of the Fabry-Perot cavity. The BMV collaboration therefore presents results in terms of the birefringence per tesla squared per pulse. At present BMV has obtained a best $\Delta n_B=5\times10^{-20}$~T$^{-2}$ per pulse and claims that a future upgrade will able to reach SNR~=~1 in the QED measurement with 1000 pulses. Given a number of pulses $N_{\rm pulses}$ with equivalent noise, the limit on $\Delta n_{\rm B}$ will scale with $1/\sqrt{N_{\rm pulses}}$. In the detection of a birefringence much smaller than $\Delta n_{\rm B}$ many pulses will be needed and therefore, to compare their method with the Q \& A or PVLAS capability, it is reasonable to rescale $\Delta n_{\rm B}$ per pulse to the average sensitivity in $\Delta n$ achievable in 1 s: $\Delta n_{\rm eff} = \Delta n_{\rm B}B^2_{\rm ext}\sqrt{T_{\rm pulse}/D_t}$. Since BMV has a small duty cycle $D_t$ their effective optical sensitivity reduces to $s_{\rm eff}=s/\sqrt{D_t}$.




It is apparent that all the experiments are still far from their goal, the chance of their eventual success relying on improvements of the optical sensitivity that, unlike hardware upgrades, cannot be given for granted, as they would require complete understanding of the noise sources limiting the sensitivity. Indeed the shot noise limited sensitivity of all the experiments, given by formula (\ref{shot}) (due to the homodyne technique adopted by BMV they have an extra factor of $1/\sqrt{2}$), is about a factor between 10 to 100 below the current achieved sensitivities. No clear explanation to this discrepancy has yet been given. Note that the best value of sensitivity in ellipticity in ellipsometers using high finesse Fabry-Perot interferometers has been reported by PVLAS in a small test ellipsometer without the implementation of a magnetic field \cite{towardsQED} of $2\times 10^{-8}~{\rm Hz}^{-1/2}$ above 5 Hz. The best reported sensitivity of a complete apparatus is by BMV with a sensitivity of $5\times10^{-8}~{\rm Hz}^{-1/2}$ over a time span of 2~ms. If in Table \ref{summary} the ÊQED ellipticity planned by each experiment is combined with the presently \emph{achieved} sensitivities, the last line in the table would become (3.2, 39, 1.5)~yr for the three experiments, respectively, instead of (12, 1.4, 8.3)~d.

\section{Conclusions}
We have presented the physics the PVLAS experiment is aiming at studying and have briefly discussed the experimental method. Noise sources are being studied on a bench-top small test apparatus in Ferrara, Italy, in view of the construction of the final apparatus with which we hope to measure for the first time the magnetic birefringence of vacuum due to vacuum fluctuations.

We have discussed the importance of using not one but two dipole magnets whose relative directions can be made perpendicular to each other in order to have a zero effect condition with the magnetic field present. This is vital to study and eliminate spurious signals generated by the rotating field. A proof of principle measurement using the Cotton-Mouton effect in oxygen gas has been done with the test apparatus.

Noise floor measurement results have been presented. A new limit on the parameter $A_e$ describing non linear electrodynamic effects in vacuum has been obtained: $A_e < 3.3\times10^{-21}$~T$^{-2}$. This value, obtained by using two small compact permanent magnets, improves the previous limit by a factor 2.

Finally, the perspectives of the three different experiments based on similar techniques active on this subject have been discussed. Ellipticity noise sources are under study by all groups so as to improve their optical sensitivity. Without such improvements the measurement of vacuum magnetic birefringence is still out of reach.

\end{document}